\documentclass[pre]{revtex4}
\usepackage{amsmath}
\usepackage{enumerate}
\usepackage{graphicx}

\begin{document}

\title{Dispersion relation of finite amplitude Alfv\'en wave 
in a relativistic electron-positron plasma}
\author{T. Hada$^a$,  S. Matsukiyo$^a$ and V. Mu\~noz$^{a,b}$}
\affiliation{$^a$ Department of Earth System Science and Technology, Kyushu
  University, Fukuoka 816-8580, Japan}
\affiliation{$^b$ Departamento de F\'{\i}sica, Facultad de Ciencias,
Universidad de Chile, Casilla 653, Santiago, Chile}
\begin{abstract}
  
  The linear dispersion relation of a finite amplitude, parallel,
  circularly polarized Alfv\'en wave in a relativistic
  electron-positron plasma is derived. In the nonrelativistic regime,
  the dispersion relation has two branches, one electromagnetic wave,
  with a low frequency cutoff at $\sqrt{1+2\omega_p^2/\Omega_p^2}$ (where
  $\omega_p=(4\pi n e^2/m)^{1/2}$ is the electron/positron plasma
  frequency), and an Alfv\'en wave, with high frequency cutoff at the
  positron gyrofrequency $\Omega_p$. 
  There
  is only one forward propagating mode for a given frequency. However,
  due to relativistic effects, there is no low frequency cutoff for
  the electromagnetic branch, and there appears a critical wave number
  above which the Alfv\'en wave ceases to exist.  This critical wave
  number is given by $ck_c/\Omega_p=a/\eta$, where
  $a=\omega_p^2/\Omega_p^2$ and 
  $\eta$ is the ratio between the Alfv\'en wave magnetic field amplitude
  and the background magnetic field. In this case, for each frequency
  in the Alfv\'en branch, two additional forward propagating modes exist
  with equal frequency.
  
  A simple numerical example is studied: by numerically solving the
  coupled system of fluid and Maxwell equations, normal incidence of a
  finite amplitude Alfv\'en wave on an interface between two
  electron-positron plasmas of different densities is considered.

\end{abstract}
\maketitle

\section{Introduction}

Electron-positron plasmas  are different from electron-ion plasmas,
because in the absence of a mass difference, there are no high or low
natural frequency scales.\cite{Tsytovich} Such plasmas are found in
pulsar magnetospheres,\cite{Curtis} models of
primitive Universe,\cite{Tajima} active galactic nuclei
jets,\cite{Wardle,Hirotani_a} and laboratory and tokamak 
plasmas.\cite{Zank,Berezhiani_a} Relativistic effects are expected to play an
important role in several of these systems.
Understanding interactions between waves and relativistic 
electron-positron plasmas is relevant
to proposed pulsar emission mechanisms \cite{Luo,Mahajan},
and may give insight into structure formation in the early
Universe \cite{Berezhiani_a}. 

Therefore, wave propagation in relativistic electron-positron plasmas
has been the subject of many studies, either in the fluid or the
kinetic treatments: linear waves \cite{Arons_a,Luo_a}, nonlinear
waves \cite{Gedalin_b,Machabeli}, and nonlinear decays
\cite{Munoz_b,Shukla_b,Matsukiyo,Munoz_m}.

In this article we deal with an Alfv\'en wave propagating along a
constant magnetic field in a pair plasma. When fully relativistic
effects are considered in the particle motion, the dispersion
relation exhibits unique features which, to our knowledge, have not
been discussed before. We then outline the numerical strategies we are
currently considering to examine the consequences of such features. 

\section{Dispersion relation}

We assume that the electron-positron plasma is described by
the following set of equations: 
\begin{gather}
\label{cont}\frac{\partial n_j}{\partial t}=-\vec{\nabla}\cdot \left(
n_j \vec v_j \right)  \ , \\
\label{force}\left(\frac{\partial }{\partial t}+\vec{v}_j\cdot\vec{\nabla}%
\right)(\gamma_j \vec{v}_j)= \frac{q_j}{m_j}\left(\vec{E}+\frac{1}{c}\vec{v%
}_j\times\vec{B}\right) \ , \\
\vec{\nabla}\cdot\vec E =4\pi\rho \ , \\
\label{faraday}\vec{\nabla}\times\vec{E}=-\frac{1}{c}\frac{\partial\vec{B}}{%
\partial t} \ , \\
\vec{\nabla}\times\vec{B}=\frac{4\pi}{c}\vec{J}+\frac{1}{c}\frac{\partial 
\vec{E}}{\partial t} \ , 
\end{gather}
\begin{gather}
\vec{J}=\sum_jq_j n_j\vec{v}_j \ , \\
\rho=\sum_jq_jn_j  \ , \\
\label{gamma}
\gamma_j=\left(1-\frac{\vec v_j\,^{2}}{c^{2}}\right)^{-1/2}  \ , 
\end{gather}
where $n_j$ is the density of each fluid, $\vec{v}_j$ is the bulk velocity
of each fluid, $\vec{E}$ and $\vec{B}$ are the electric and magnetic fields,
respectively, $\vec J$ is the total
current,   $m$ is 
the particle mass, and $c$ is the speed of light. $j=p$ for positrons,
and $j=e$ for electrons.

We assume that a circularly polarized Alfv\'en wave propagates
along the $z$-axis, as well as the existence of a constant magnetic field in
the same direction, $B_{0}\hat z$. The wave  electric and magnetic
fields are given by: 
\begin{equation}
\label{vecB}
\vec B = B [ \hat x \cos (k z - \omega t) + \hat y \sin (k z -
\omega t) ] \ ,
\end{equation}
\begin{equation}
\label{vecE}
\vec E = \frac{\omega}{ck} B
[ \hat x \sin (k z - \omega t) - \hat y \cos (k z -
\omega t) ]  \ . 
\end{equation}
Introducing \eqref{vecB} and \eqref{vecE} in the fluid equations,
the transverse velocity for each species is obtained as \cite{Matsukiyo}
\begin{equation}
  \label{v}
  \frac{v_j}c =
  \frac\omega{ck}\frac{\Omega_j}{\gamma_j\omega-\Omega_j} \eta \ ,
\end{equation}
where $\Omega_j = q_j B_0/m_jc$ is the cyclotron frequency, and
$\eta=B/B_0$. The dispersion relation for the Alfv\'en wave is
\begin{equation}
  \label{disprel}
  \frac{c^2k^2}{\omega^2} = 1-\sum_j
  \frac{\omega_j^2}{\omega(\gamma_j\omega-\Omega_j)} \ ,
\end{equation}
where $\omega_j = (4\pi n_{0j}q_j^2/m_j)^{1/2}$ is the plasma frequency
and $n_{0j}$ is the rest density of species $j$.

We want to numerically solve the dispersion relation, which is
equivalent to simultaneously solving the set of equations
\eqref{gamma}, \eqref{v} and \eqref{disprel}.
For an electron positron plasma, $\omega_p=\omega_e$. We define 
\begin{equation}
  \label{normalization}
  x=\frac \omega{\Omega_p} \ , \quad y = \frac{ck}{\Omega_p} \ , \quad a =
  \frac{\omega_p^2}{\Omega_p^2} \ , \quad u = \frac vc \ .
\end{equation}
Normalized equations \eqref{gamma}, \eqref{v} and \eqref{disprel}, are
\begin{gather}
  \label{normalized_gamma}
  \gamma_j = (1-u_j^2)^{-1/2} \ , \\
\label{normalized_v}
u_j = \frac xy \eta \frac{\pm 1}{x\gamma_j \mp 1} \ , \\
\label{normalized_disprel}
y^2 = x^2 - \frac{ax}{x\gamma_p-1} - \frac
{ax}{x\gamma_e+1} \ ,
\end{gather}
where the upper (lower) sign in \eqref{normalized_v} is for positrons
(electrons). 


Eliminating $u_j$, the following equation is obtained for $\gamma$:
\begin{equation}
  \label{polynomial_gamma}
  \gamma_j^4 + \gamma_j^3\left(\mp\frac2x\right) + \gamma_j^2
  \left(-1+ \frac 1{x^2} - \frac{\eta^2}{y^2}\right) + \gamma_j
  \left(\pm \frac 2x\right) - \frac 1{x^2} = 0 \ . 
\end{equation}

\begin{figure}[h]
  \centering
  \includegraphics[width=7cm]{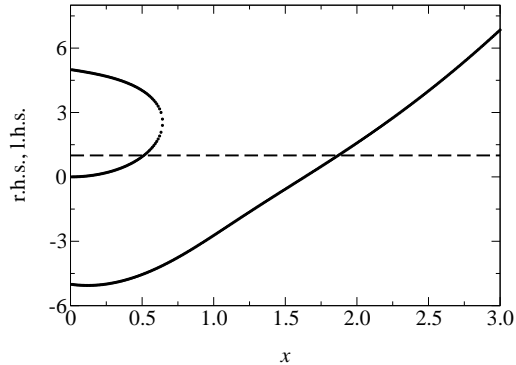}
  \caption{Graphical representation of \eqref{normalized_disprel}, for
    $a=1$, $y=ck/\Omega_p=1$, $\eta=0.2$. Full circles: right hand size
    of \eqref{normalized_disprel}; dashed line: left hand size of
    \eqref{normalized_disprel}}.  
  \label{rhs}
\end{figure}

For a given $y$, from \eqref{polynomial_gamma} $\gamma_j$ is calculated
as a function of $x$, and then the right hand size of
Eq.~\eqref{normalized_disprel} can be plotted. This is shown in
Fig.~\ref{rhs}.  Since $y$ is given, we plot the left side of
Eq.~\eqref{normalized_disprel} on the same graph, and the intersection
points of the curves are the roots of the dispersion relation. The
curve on the right corresponds to the light wave branch of the
dispersion relation, and the curve on the left corresponds to the
Alfv\'en branch. It is interesting to note that when $y>y_c=a/\eta$
the Alfv\'en wave ceases to exist.

\begin{figure}[h]
  \centering 
  \includegraphics[width=5.5cm]{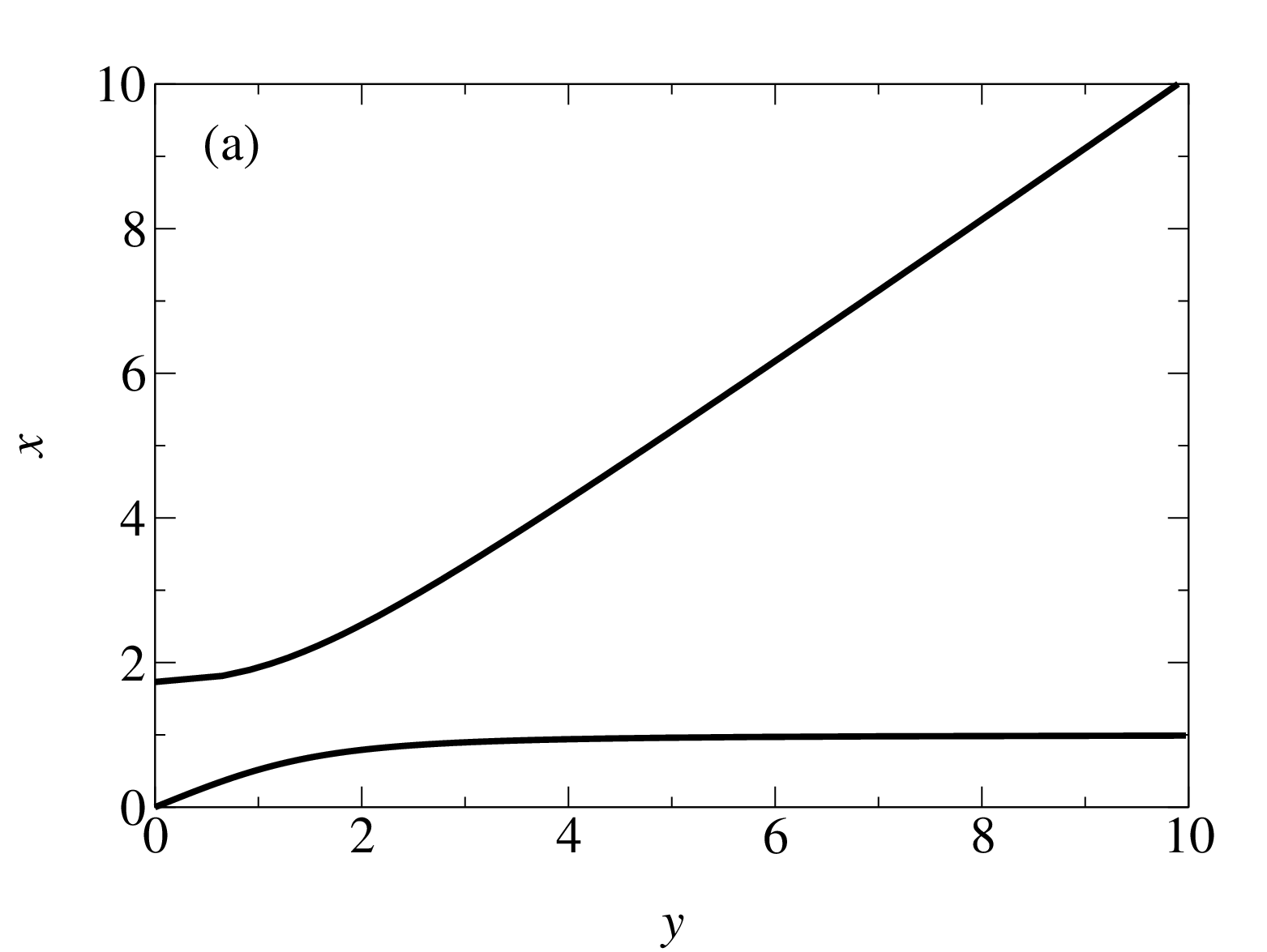}
  \includegraphics[width=5.5cm]{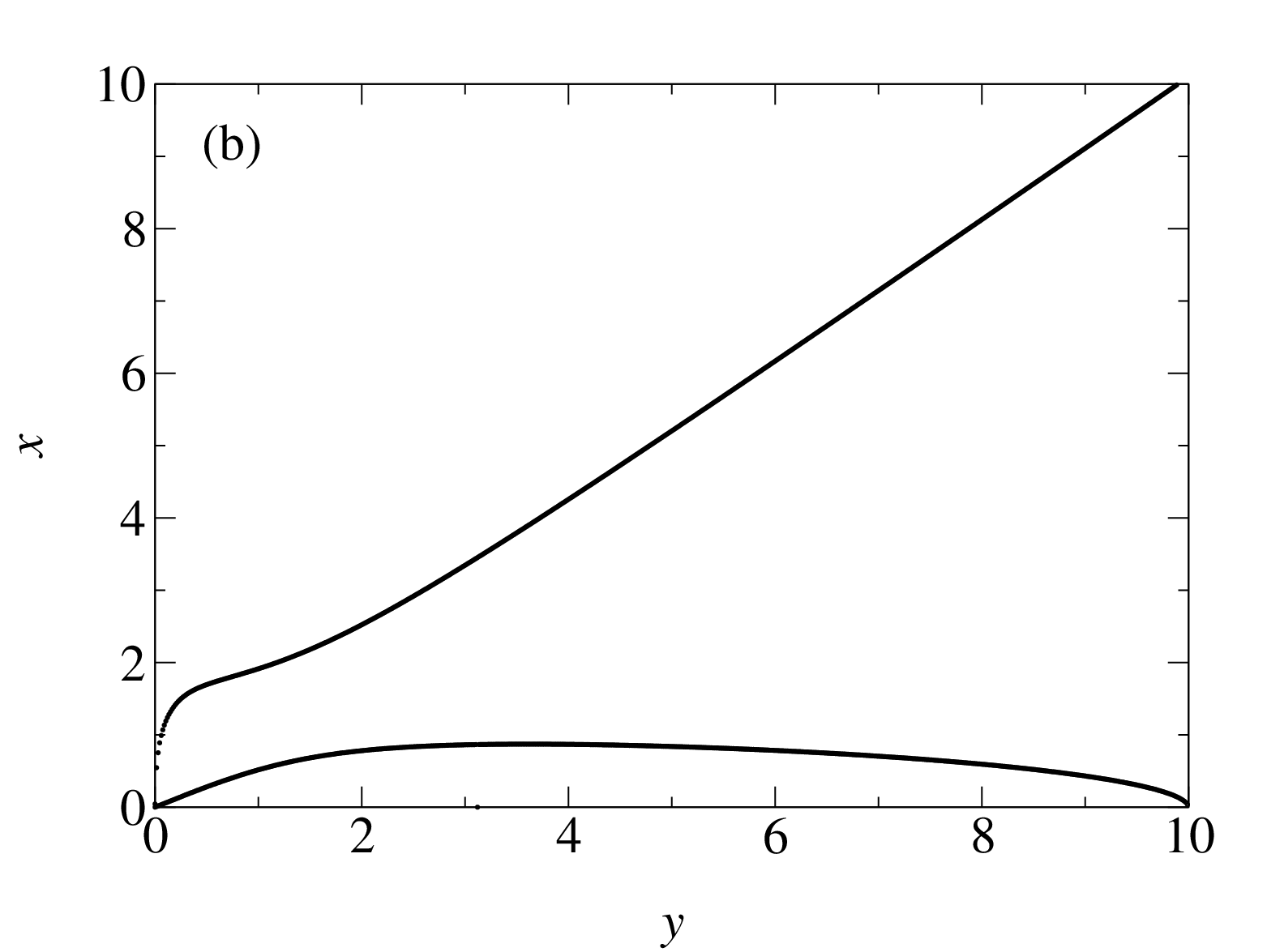}
  \includegraphics[width=5.5cm]{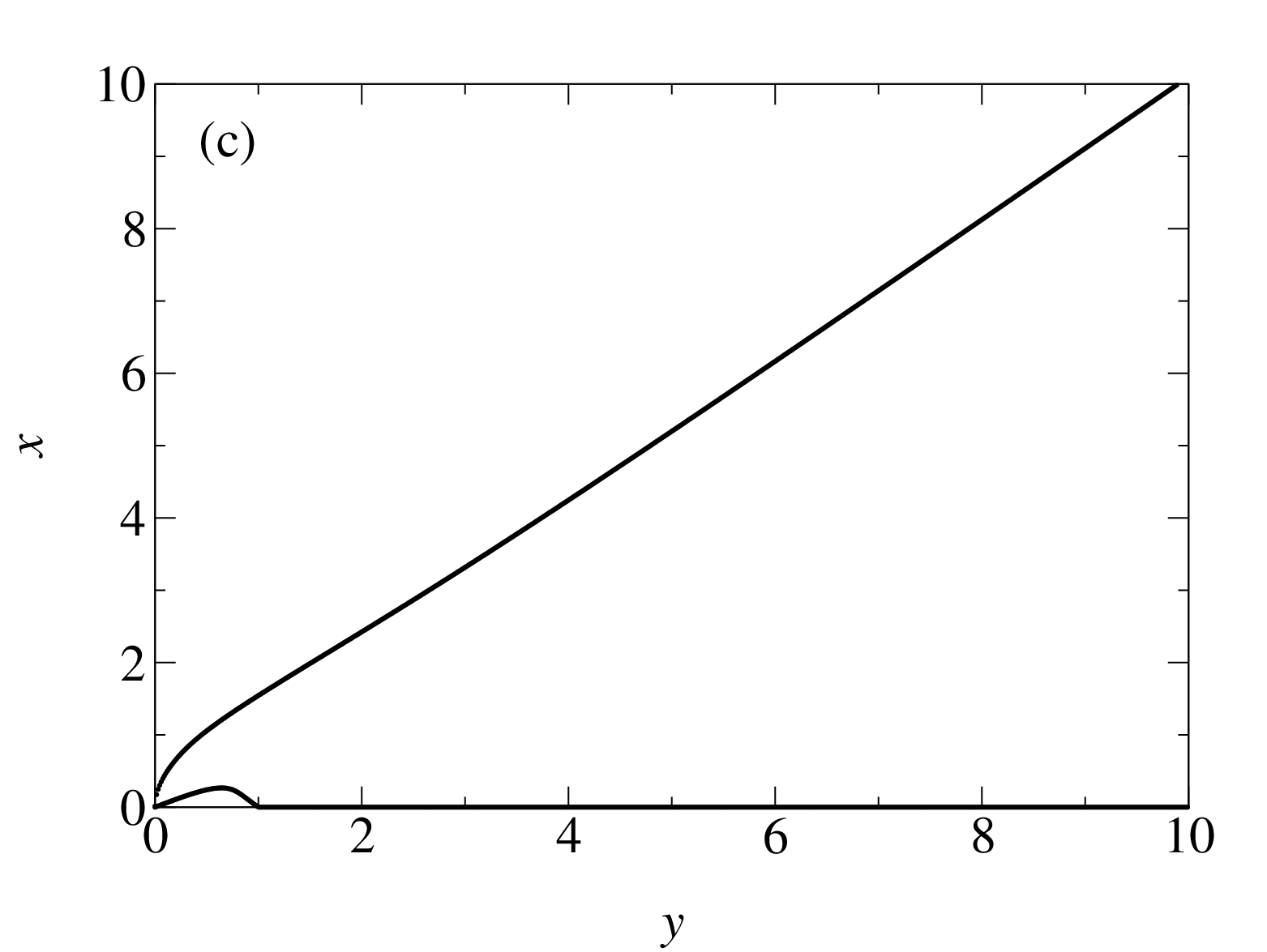}
  \caption{Dispersion relation \eqref{normalized_disprel} for (a)
    $a=1$, $\eta=0$; (b)
    $a=1$, $\eta=0.1$; (c) $a=1$, $\eta=1$. Since the relevant
    parameter is $y_c=a/\eta$, Figs.~(b) and (c) are reobtained if
    $\eta=0.1$ is constant, whereas $a$ is decreased from $a=1$ to $a=0.1$.}
  \label{fig_disprel}
\end{figure}

Thus dispersion relation \eqref{normalized_disprel} can be
solved. The result is shown in Fig.~\ref{fig_disprel}. 
For $\eta=0$ [Fig.~\ref{fig_disprel}(a)], the nonrelativistic result
is recovered.  There are two branches, one corresponds to the light
wave, with a cutoff at $\omega/\Omega_p=\sqrt{1+2a}$. 
The other one is the
Alfv\'en branch, which has a resonance at the positron gyrofrequency.
For $\eta=0.1$ [Fig.~\ref{fig_disprel}(b)], however, the light wave
branch has no low frequency cutoff, and the Alfv\'en branch stops at
$y_c=a/\eta$. For even higher values of the wave amplitude $\eta$ [see
Fig.~\ref{fig_disprel}(c) for $\eta=1$], the Alfv\'en branch exists
for a very short wavenumber range. The Alfv\'en branch is also
constrained to a shorter frequency range. This can also be seen in
Fig.~\ref{rhs}. From
Eqs.~\eqref{normalized_gamma}--\eqref{normalized_disprel} it can be
shown that the Alfv\'en branch in Fig.~\ref{rhs} has a frequency
cutoff at the critical 
frequency
\begin{equation}
  \label{xc}
  x_c=\frac{\omega_c}{\Omega_p} = \left[1+\left(\frac\eta
      y\right)^{2/3}\right]^{-3/2} \ .
\end{equation}
This cutoff depends on the wave amplitude $\eta$, unlike the
nonrelativistic result, where $\omega_c=\Omega_p$. 

\section{Outline of the numerical study}

Fig.~\ref{fig_disprel} shows that, for a given frequency $\omega$, the
existence of Alfv\'en waves and the number of wave modes 
depend on the physical parameters of the plasma. This poses a number
of new questions. For instance, we could consider the problem
of normal incidence of a plane wave on a density interface. In the
usual problem of a light wave in a dielectric medium, only one mode
exists for a given frequency. Given only the boundary conditions that
the fields must satisfy, it is possible to completely solve the
problem and obtain explicit solutions for the field amplitude at both
sides of the interface. However, now several modes exist for a given
frequency. Moreover, if an Alfv\'en wave exists with a given frequency
in one side of the interface, then if the density is small enough on
the other side it could not exist. Does it become an evanescent
wave? Is it converted to other modes? 

In order to investigate this we are studying the system by means of a
fluid simulation, and by direct time integration of the evolution
equations. 

For the fluid simulation, we notice that the system equations lead to
the following wave equations for the 
electromagnetic field:
\begin{align}
  \label{waveB}
  \left( \frac 1{c^2} \frac{\partial^2}{\partial t^2} -
    \nabla^2\right) \vec B &= \frac{4\pi}c e n_p \vec\nabla\times(\vec
  v_p-\vec v_e) \ ,\\
  \label{waveE}
  \left( \frac 1{c^2} \frac{\partial^2}{\partial t^2} -
    \nabla^2\right) \vec E &= -\frac{4\pi}c e n_p
  \frac{\partial}{\partial t}(\vec
  v_p-\vec v_e) \ . 
\end{align}

The wave equations for the fields \eqref{waveB} and
\eqref{waveE}, and the momentum equation \eqref{force}, are
discretized using
time-centered and space-centered finite differences. 

The second approach we are developing is time integration of the
evolution equations \eqref{cont}-\eqref{gamma} by means of the
rationalized Runge-Kutta method.

Both procedures are under development/testing. A
particularly simple problem is that of a pure Alfv\'en mode
propagating in a relativistic plasma. Since in a pair plasma there is
no harmonic generation, all oscillatory fields and velocities are
purely transverse. We are currently working to obtain satisfactory
results for this problem, in order to later consider incidence on a
normal interface, where longitudinal oscillations may not be
neglected. 

\section{Conclusions}

Propagation of a finite amplitude Alfv\'en wave in an
electron-positron plasma has been studied. Full relativistic effects
on the particle velocities in the wave field have been considered. The
dispersion relation for propagation along a constant magnetic field
has been obtained and numerically solved. Several features are
different to the usual nonrelativistic result
[Fig.~\ref{fig_disprel}(a)]. 
In the nonrelativistic case, there
are two branches, an Alfv\'en wave and a light wave. The light wave
has a low frequency cutoff at the frequency
$\omega/\Omega_p=\sqrt{1+2a}$, and the Alfv\'en wave has a high frequency
cutoff at the positron gyrofrequency, $\omega=\Omega_p$. However, in
the relativistic case the light wave branch does not have a low
frequency cutoff. As to the other branch, there are two Alfv\'en wave
modes for any given frequency, and there is a high wavenumber cutoff at $c
k_c/\Omega_p =a/\eta$. Besides, the high frequency cutoff for the
Alfv\'en wave depends on the plasma parameters. For large
 amplitude wave or small plasma density, the Alfv\'en branch is
constrained to a small region of frequencies and wavenumbers [see
Fig.~\ref{fig_disprel}(c)]. 

We intend to investigate the behavior of the system when a
relativistic Alfv\'en wave of frequency $\omega$ propagates through a
pair plasma, and then finds a region where plasma parameters are such
that such wave cannot propagate. The existence of more normal modes
than in the usual nonrelativistic problem make this a nontrivial
matter. As a first example, we are considering the simple problem of
normal incidence of an Alfv\'en wave on a planar density
discontinuity. Two approaches are being developed, namely a fluid
simulation, and time integration of the evolution equations. Work on
this is in progress, and we expect to present results elsewhere. 

It is worth noting that for waves in the high-wavenumber end of the
Alfv\'en branch in Fig.~\ref{fig_disprel}, particles have large
Lorentz factors, and therefore synchrotron radiation emission will be
important. Thus, the results presented here may be further modified
when additional effects such like synchrotron loss are taken into
account.


\begin{thebibliography}{10}

\bibitem{Tsytovich}
V.~Tsytovich and C.~B. Wharton, Comments Plasma Phys.\ Controlled Fusion {\bf
  4}, 91 (1978).

\bibitem{Curtis}
M.~F. Curtis, {\em The Theory of Neutron Stars Magnetospheres\/} (University of
  Chicago Press, Chicago, 1991).

\bibitem{Tajima}
T.~Tajima and T.~Taniuti, Phys.\ Rev.\ A {\bf 42}, 3587 (1990).

\bibitem{Wardle}
J.~F.~C. Wardle, D.~C. Homan, R.~Ojha, and D.~H. Roberts, Nature {\bf 395}, 457
  (1998).

\bibitem{Hirotani_a}
K.~Hirotani, S.~Iguchi, M.~Kimura, and K.~Wajima, Astrophys.\ J. {\bf 545}, 100
  (2000).

\bibitem{Zank}
G.~P. Zank and R.~G. Greaves, Phys.\ Rev.\ E {\bf 51}, 6079 (1995).

\bibitem{Berezhiani_a}
V.~I. Berezhiani and S.~M. Mahajan, Phys.\ Rev.\ Lett. {\bf 73}, 1110 (1994).

\bibitem{Luo}
Q.~Luo and D.~B. Melrose, Mon.\ Not.\ R. Astron.\ Soc. {\bf 258}, 616 (1992).

\bibitem{Mahajan}
S.~M. Mahajan, Astrophys.\ J. Lett. {\bf 479}, L129 (1997).

\bibitem{Arons_a}
J.~Arons and J.~J. Barnard, Astrophys.\ J. {\bf 302}, 120 (1986).

\bibitem{Luo_a}
Q.~Luo, D.~B. Melrose, and D.~Fussell, Phys.\ Rev.\ E {\bf 66}, 026405 (2002).

\bibitem{Gedalin_b}
M.~Gedalin, D.~B. Melrose, and E.~Gruman, Phys.\ Rev.\ E {\bf 57}, 3399 (1998).

\bibitem{Machabeli}
G.~Z. Machabeli, S.~V. Vladimirov, and D.~B. Melrose, Phys.\ Rev.\ E {\bf 59},
  4552 (1999).

\bibitem{Matsukiyo}
S.~Matsukiyo and T.~Hada, Phys.\ Rev.\ E {\bf 67}, 046406 (2003).

\bibitem{Munoz_b}
V.~Mu\mbox{\~n}oz and L.~Gomberoff, Phys.\ Rev.\ E {\bf 57}, 994 (1998).

\bibitem{Shukla_b}
P.~K. Shukla and L.~Stenflo, Phys.\ Plasmas {\bf 7}, 2726 (2000).

\bibitem{Munoz_m}
V.~Mu{\~n}oz, Phys.\ Plasmas {\bf 11}, 3497 (2004).

\end{thebibliography}

\end{document}